\documentclass[usenatbib,usegraphicx]{mn2e}
\usepackage{bm}
\usepackage{times}

\title[Tidal mass loss in star clusters]
{Tidal mass loss in star clusters and treatment of escapers in Fokker-Planck models}
\author[K. Takahashi and H. Baumgardt]{K. Takahashi$^{1}$\thanks{E-mail:
tkoji@sit.ac.jp (KT)} and H. Baumgardt$^{2}$\\
$^{1}$Department of Informational Society Studies, Faculty of Human and Social Studies, Saitama Institute of Technology, \\
1690 Fusaiji, Fukaya, Saitama 369-0293, Japan\\
$^{2}$School of Mathematics and Physics, University of Queensland, Brisbane, QLD 4072, Australia
}

\begin{document}

\date{Accepted 201x xx xx. Received 2011 xx xx; in original form 2011 xx xx}

\pagerange{\pageref{firstpage}--\pageref{lastpage}} \pubyear{201x}

\maketitle

\label{firstpage}

\begin{abstract}
This paper presents a new scheme to treat escaping stars in the orbit-averaged Fokker-Planck models
of globular star clusters in a galactic tidal field.
The existence of a large number of potential escapers, which have energies above the escape energy but are still 
within the tidal radius, is taken into account in the models.
The models allow potential escapers to experience gravitational scatterings before they leave clusters
and thus some of them may lose enough energy to be bound again.
It is shown that the mass evolution of the Fokker-Planck models are in good agreement with 
that of $N$-body models including the full tidal-force field.
The mass-loss time does not simply scale with the relaxation time due to the existence of potential escapers;
it increases with the number of stars more slowly than the relaxation time,
though it tends to be proportional to the relaxation time in the limit of a weak tidal field. 
The Fokker-Planck models include two parameters, the coefficient $\gamma$ in the Coulomb logarithm $\ln (\gamma N)$
and the coefficient $\nu_{\rm e}$ controlling the efficiency of the mass loss.
The values of these parameters are determined by comparing the Fokker-Planck models with the $N$-body models.
It is found that the parameter set $(\gamma, \nu_{\rm e})=(0.11, 7)$ works well 
for both single-mass and multi-mass clusters, 
but that the parameter set $(\gamma, \nu_{\rm e})=(0.02, 40)$ is another possible choice for multi-mass clusters.
\end{abstract}


\section{Introduction} \label{sec:intro}

The numerical integration scheme of the orbit-averaged Fokker-Planck (FP) equation developed by \citet{c79} has been one of 
the most useful tools for simulating the dynamical evolution of globular star clusters.
In addition to two-body relaxation, many physical processes  have been incorporated into FP models to achieve
realistic modelling of the globular cluster evolution; 
these processes include tidal cutoff, binary heating, disc and bulge shocks, mass loss via stellar evolution, etc.
(see \citealt*{skt08} for a recent example of detailed FP modelling).

In this paper we consider the dynamical evolution of globular clusters in a steady galactic tidal field.
Our main purpose is to investigate what boundary condition can give a better description 
of escape of stars from clusters in the tidal field.
This study has been motivated by the studies by \citet{fh00} and \citet{b01}.

\citet{fh00} found that a large fraction of stars with energies above the escape energy (i.e. potential escapers)
take much longer escape time than the dynamical time.
Until their study it had been generally thought that the escape time-scale is of the order of the dynamical time
and that the mass-loss times of the clusters essentially scale with the relaxation time,
which is much longer than the dynamical time.
The findings of \citet{fh00} indicate that this simple scaling may be spoiled by potential escapers 
with long escape times. 

In fact \citet{b01} performed $N$-body simulations and showed that the mass-loss times (lifetimes) of clusters 
do not scale with the relaxation time $t_{\rm rh}$ but scale with $t_{\rm rh}^{3/4}$.
He concluded that the reason is that some of potential escapers are scattered back to lower energies before they
leave the cluster.
More recently \citet{tf05} showed that the dependence on the relaxation time changes with the strength of the tidal filed.
These two studies have revealed that
the behavior of potential escapers greatly influences the rate of mass loss from clusters in the tidal field.

The effects of long escape times and re-scattering of potential escapers have never been considered 
in previous FP models in the literature,
but it was assumed that escapers leave a cluster on the dynamical time-scale,
as is described in detail in Section~\ref{sec:FP}.
Since the effect of the galactic tidal field is essentially important to the cluster evolution,
it is necessary to find a way to include the effect into FP models as precisely as possible.

We should mention that
Takahashi \& Portegies Zwart (1998, 2000) compared FP and $N$-body models of star clusters in the tidal field
and found good agreement between these two theoretical models over a wide range of initial conditions.
They showed that the use of anisotropic FP models with the apocentre escape criterion \citep*{tli97}
and the dynamical-time removal of escapers \citep{lo87}
is necessary to obtain such good agreement.
However, note that in their $N$-body models 
the tidal force field is not included but the tidal cutoff is applied.
\citet{tpz00} confirmed that the difference between
tidal cutoff and self-consistent tidal field $N$-body models is small
for a particular set of initial conditions,
but did not do systematic investigations on this problem.

In this study we have devised a new scheme to treat escapers in FP models.
The scheme defines a region of potential escapers in phase space and allows them to be scattered again.
Comparing the results of FP models calculated with the new scheme with the results of $N$-body models,
we examine the accuracy of the FP models.

\section{Fokker--Planck models of star clusters in a steady tidal field} \label{sec:FP}

\subsection{Basic assumptions}

The orbit-averaged FP equation is derived under the assumption of spherical symmetry of star clusters \citep{c79}.
Therefore the tidal field, which is not spherically symmetric, cannot be directly incorporated into 
orbit-averaged FP models.
In FP models the effect of the tidal field is taken into account 
by imposing a tidal cutoff radius $r_{\rm t}$ on the cluster,
which is treated as an isolated system in other respects.
Under these assumptions the distribution function $f$ of stars at time $t$ depends only on
the energy of a star per unit mass, $E$, and the angular momentum per unit mass, $J$.

\subsection{Classical treatments of escapers}

First we summarise classical treatments of escapers used in FP models of previous studies.

\subsubsection{Escape criteria in phase space} \label{sec:esc_crit}

In previous studies,
two kinds of criteria were adopted to define an escape region in $(E,J)$-space:
\begin{enumerate}
\item Energy criterion
\begin{equation}
E > E_{\rm t} \equiv -\frac{GM}{r_{\rm t}} \label{eq:eng-crit},
\end{equation}
\item Apocentre criterion
\begin{equation}
r_{\rm a}(E,J) > r_{\rm t} \label{eq:apc-crit},
\end{equation}
\end{enumerate}
where $M$ is the cluster mass and $r_{\rm a}(E,J)$ is the apocentre radius of a star having energy $E$ and angular momentum $J$.
It is assumed that a star is destined to escape once it enters into the escape region.

The apocentre criterion \citep{tli97} is considered to be more realistic, 
at least as long as the tidal field is modelled as a radial cut-off,
and in fact gives better agreement between FP and $N$-body models \citep{tpz98, tpz00}.
For isotropic FP models, where the distribution function does not depend on $J$,
 only the energy criterion can be applied (e.g. \citealt{lo87}).

\subsubsection{Removal of escapers}

In previous studies stars in the escape region are assumed to leave the cluster inevitably,
as mentioned above.
It is also assumed that 
the time required for this travel is of the order of the dynamical time at the tidal radius.
Considering this travel time, \citet{lo87} applied the following equation 
to the distribution function $f$ in the escape region:
\begin{equation}
     \frac{\partial f}{\partial t}
     = - \nu_{\rm e} f \left[ 1- \left( \frac{E}{E_{\rm t}} \right)^3 \right]^{1/2} 
       /t_{\rm tid},
\label{eq:lo}
\end{equation}
where $\nu_{\rm e}$ is a dimensionless constant determining the efficiency of escape
(see also \citealt*{lfr91}).
The time-scale $t_{\rm tid}$ is an orbital time-scale at the tidal radius defined by
\begin{equation}
t_{\rm tid}=\frac{2\pi}{\sqrt{(4\pi/3){G \rho_{\rm t}}}},
\end{equation}
where $\rho_{\rm t}$ is the mean mass density within the tidal radius.

Since the dynamical time is generally much smaller than the relaxation time
in globular clusters,  we may assume that escapers leave the cluster immediately after they
enter into the escape region, when we are interested only in the evolution on the relaxation time-scale.
This assumption leads to the boundary condition
\begin{equation}
f=0
\end{equation}
on the tidal boundary (e.g. \citealt{cw90}).

\subsection{A new treatment of escapers}\label{sec:nt}

The boundary condition of equation (\ref{eq:lo}) takes account of the fact that 
stars satisfying the escape criterion, i.e. potential escapers, need time to actually leave the cluster.
However the effect of re-scattering of potential escapers is not considered there.
Here we propose a new scheme in which the re-scattering effect is taken into account.

First we summarise basic assumptions and equations.
Suppose that the cluster is on a circular orbit,
with radius $R_{\rm G}$ and angular velocity $\omega$, round the centre of a spherical galaxy. 
We consider the motion of a star in the rotating coordinate system moving with the cluster;
the origin is at the cluster centre, the $x$-axis points to the galactic centre,
and the $y$-axis is in the cluster orbital plane.
If the cluster and the galaxy are treated as point masses $M$ and $M_{\rm G}$ ($\gg M$)
and the size of the cluster is much smaller than  $R_{\rm G}$,
there exists a conserved quantity known as the Jacobi integral given by
\begin{equation}
 E_{\rm J}=\frac{v^2}{2}-\frac{GM}{r}-\frac12 \omega^2 (3x^2-z^2) \label{eq:ej}
\end{equation}
(cf. \citealt{s87}, Chapt. 5).
Here $v$ is the velocity of the star measured in the rotating frame,
$r$ is the distance from the star to the cluster centre,
and the angular velocity $\omega$ is given by
\begin{equation}
 \omega=\sqrt{\frac{GM_{\rm G}}{R_{\rm G}^3}} .
\end{equation}
The third term on the right-side in equation (\ref{eq:ej}) is a combination of the centrifugal and tidal potentials. 

The effective potential is defined as
\begin{equation}
 \phi_{\rm eff}(x,y,z)=-\frac{GM}{r}-\frac12 \omega^2 (3x^2-z^2).
\end{equation}
A contour plot of $\phi_{\rm eff}$ is shown, e.g., in Fig. 5.1 of \citet{s87}.
The effective potential has the saddle points at $(\pm x_{\rm e},0,0)$, where
\begin{equation}
x_{\rm e}=\left( \frac{M}{3M_{\rm G}}\right)^{1/3} R_{\rm G}
\end{equation}
and
\begin{equation}
  \phi_{\rm eff}(\pm x_{\rm e},0,0)  =-\frac32 \frac{GM}{x_{\rm e}}.  \label{eq:pxe}
\end{equation}
The equipotential surface passing through these saddle points intersects with the $y$-axis at $y=\pm y_{\rm e}$, where
\begin{equation}
 y_{\rm e}=\frac23 x_{\rm e}.  \label{eq:ye}
\end{equation}
The necessary condition for escape of a star from the cluster
is given by
\begin{equation}
E_{\rm J} > E_{\rm J,crit} \equiv -\frac32 \frac{GM}{x_{\rm e}}.  \label{eq:ejcrit}
\end{equation}
Note that equations (\ref{eq:pxe}), (\ref{eq:ye}), and (\ref{eq:ejcrit}) are valid 
for any spherical galactic potential.
\citet{fh00} found that the time-scale for escape of stars with $E_{\rm J}>E_{\rm J,crit}$ 
varies as
\begin{equation}
 t_{\rm e} \propto ( E_{\rm J}-E_{\rm J,crit})^{-2}. \label{eq:fhte}
\end{equation}

With this relation in mind we have devised a new scheme to follow the evolution of 
potential escapers.
In this scheme the evolution of the distribution function $f$ for potential escapers is described by
\begin{equation}
\frac{\partial f}{\partial t} =
\left( \frac{\partial f}{\partial t} \right)_{\rm coll}
-\frac{f}{t_{\rm e}(E)},  \label{eq:fp_pe}
\end{equation}
where the first term on the right-side is the FP collision term
and the second term represents mass loss due to escape.
Here the escape time-scale $t_{\rm e}$ is given by
\begin{equation}
\frac{1}{t_{\rm e}(E)} = 
  \frac{\nu_{\rm e}}{t_{\rm tid}} \left( 1-\frac{E}{E_{\rm crit}} \right)^2, \label{eq:te}
\end{equation}
where $\nu_{\rm e}$ is a dimensionless numerical constant.
It should be noted that energy $E$, not the Jacobi integral $E_{\rm J}$, is used in equations (\ref{eq:fp_pe}) and (\ref{eq:te}).
Energy $E$ does not include the centrifugal and tidal potentials.
Despite this difference, we use the same critical value of energy
\begin{equation}
 E_{\rm crit} = -\frac32 \frac{GM}{r_{\rm t}}, \label{eq:ecrit}
\end{equation}
where the tidal radius $r_{\rm t}$ is identified with $x_{\rm e}$.
One might think that
using equation (\ref{eq:te}) with equation (\ref{eq:ecrit})
is too crude an approximation,
but it brings good agreement between FP and $N$-body models as is shown in Section~\ref{sec:results}.

The most important difference between equations (\ref{eq:lo}) and (\ref{eq:fp_pe})
is that the latter includes the collision term.
Thus equation (\ref{eq:fp_pe}) allows potential escapers to be scattered back to lower energies.
The effect of mass loss is included in both equations in a similar way,
though the functional forms of the escape time-scale $t_{\rm e}$ are different.

In this new treatment of the tidal field,
the escape criteria described in Section~\ref{sec:esc_crit}
are modified as follows:
\begin{enumerate}
\item Energy criterion
\begin{equation}
E > E_{\rm crit} = -\frac32 \frac{GM}{r_{\rm t}} \label{eq:eng-crit_pe},
\end{equation}
\item Apocentre criterion 
\begin{equation}
r_{\rm a}(E,J) > \frac23 r_{\rm t} \label{eq:apc-crit_pe}.
\end{equation}
\end{enumerate}
Note that $\phi_{\rm eff}(0,\pm 2r_{\rm t}/3,0) =\phi(0,\pm 2r_{\rm t}/3,0)=-3GM/2r_{\rm t}$.
Equation (\ref{eq:fp_pe}) is applied only in the region where an adopted criterion is satisfied.

\subsection{The Fokker-Planck code}

The FP code used in the present study is essentially the same as 
that used by \citet{tpz00},
but adopts the new scheme for treating escapers described above.
The code calculates the evolution of the distribution function $f(E,J,t)$.
Unlike \citet{tpz00}, stellar evolution is not considered in the models presented in this paper.
Instead the effect of heating by three-body binaries is considered in the manner described in \citet{t97}.

For all the models presented in the present paper,
201 energy mesh points, 51 angular-momentum mesh points, and 101 radial mesh points are used.
The meshes are constructed as described in \citet{t95}.
When calculating the evolution of multi-mass clusters, 
10 discrete mass-components are used to represent a continuous mass function.

Our FP models have two free parameters: one is $\nu_{\rm e}$ in equation~(\ref{eq:te}) 
and the other is $\gamma$ in the Coulomb logarithm $\ln (\gamma N)$ appearing in the FP collision term.
How the value of $\nu_{\rm e}$ is determined is described in Section~\ref{sec:results}.
We set $\gamma=0.11$ \citep{gh94a} in most of our runs 
and $\gamma=0.02$ \citep{gh96} in a part of runs for multi-mass clusters.

\section{Results} \label{sec:results}

\subsection{Comparison with $N$-body models: single-mass clusters}

First we compare FP models with the full tidal field
models of \citet{b01} and additional $N$-body runs performed for this comparison.
All the model clusters are composed of equal-mass stars
and move on circular orbits round a point-mass galaxy.
The initial distribution of stars is given by King models \citep{k66}.

Results are presented in $N$-body units, where the initial total mass and energy of a cluster
are equal to 1 and $-0.25$, respectively, and the gravitational constant $G=1$.
The same units are used throughout this paper.

Here we will refer to FP models with the boundary condition of equation (\ref{eq:fp_pe}) as ``FPf'' models,
which aim to model clusters in a self-consistent {\it full tidal field}.
FP models with equation (\ref{eq:lo}) will be called ``FPd'' models,
where stars beyond the tidal cutoff radius are removed on the {\it dynamical time-scale}.

\begin{figure}
\begin{center}
\includegraphics[width=84mm]{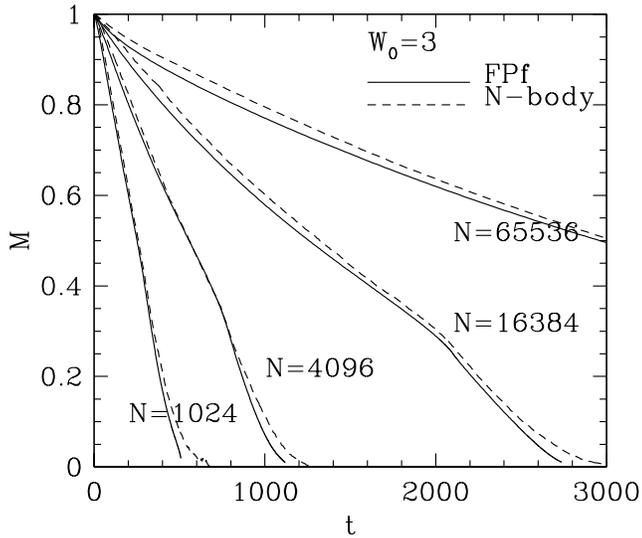}
\end{center}
\caption{Evolution of the cluster mass.
The solid lines represent FPf models,
and the dashed lines represent $N$-body models.
The initial models are $W_0=3$ King models with the number of stars $N=1024$, 4096, 16384 and 65536.
}\label{fig:massW3}
\end{figure}

Fig.~\ref{fig:massW3} compares FPf and $N$-body models concerning
the evolution of the total mass of bound stars.
The initial models are $W_0=3$ King models with the number of stars $N=1024$, 4096, 16384 and 65536.
The new treatment of escapers described by equation (\ref{eq:fp_pe})
with the apocentre criterion of equation (\ref{eq:apc-crit_pe}) is employed in the FPf models.
The agreement between the FPf and $N$-body models is good in all the cases.
In fact the value of the parameter $\nu_{\rm e}$ in equation (\ref{eq:fp_pe})
has been determined so that good agreement is obtained by performing test runs with
different values of $\nu_{\rm e}$ as was done by \citet{tpz00}.
We have finally chosen the value of $\nu_{\rm e}=7$.
All the FPf models shown in Fig.~\ref{fig:massW3} are calculated with this value.

\begin{figure}
\begin{center}
\includegraphics[width=84mm]{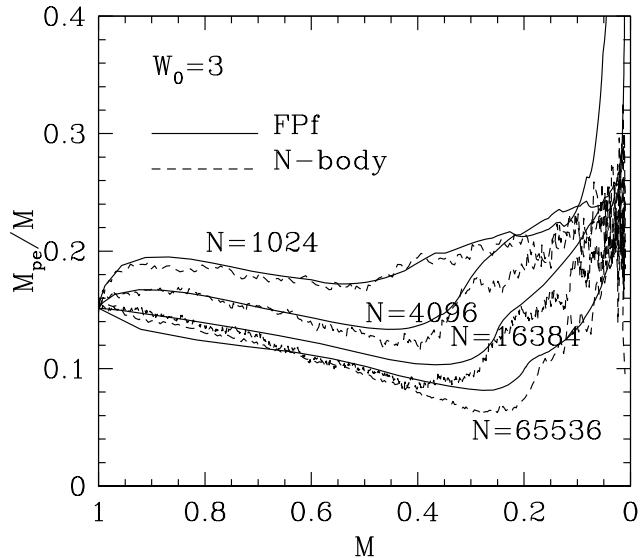}
\end{center}
\caption{Evolution of 
the ratio of the mass of potential escapers $M_{\rm pe}$ 
to the total cluster mass $M$.
The ratio is plotted as a function of the cluster mass at each instance.
}\label{fig:mpeW3}
\end{figure}

Fig.~\ref{fig:mpeW3} shows the evolution of the ratio of the mass of potential escapers $M_{\rm pe}$ 
to the total cluster mass $M$ for the runs shown in Fig.~\ref{fig:massW3}.
The agreement between the FPf and $N$-body models is fairly good 
also in this comparison.
Note that here $M_{\rm pe}$ for the FPf models is defined as the mass of stars with $E>E_{\rm crit}$,
although the apocentre criterion is used in the simulations.
The mass of stars satisfying the apocentre criterion is smaller than 
that of stars with $E>E_{\rm crit}$, but shows a similar trend in time variation.

\begin{figure}
\begin{center}
\includegraphics[width=84mm]{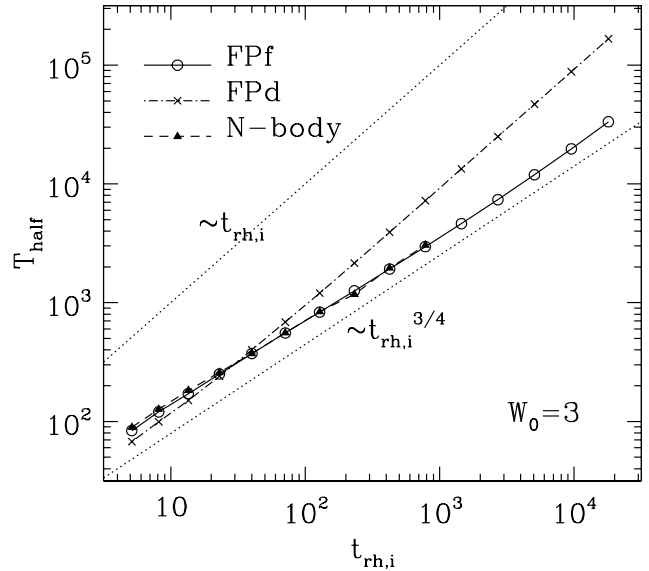}
\end{center}
\caption{Half-mass time $T_{\rm half}$ as a function of the initial half-mass relaxation time $t_{\rm rh,i}$
for the initial conditions of $W_0=3$ King models.
Two types of FP models, FPf and FPd models (see text), are shown by
the circles and crosses, respectively,
and $N$-body models are shown by the triangles.
The dotted lines represent scalings proportional to $t_{\rm rh,i}$ and $t_{\rm rh,i}^{3/4}$
(they are arbitrarily shifted in a vertical direction).
}\label{fig:thalfW3}
\end{figure}

\begin{table}
\caption{Half-mass times $T_{\rm half}$ given by $N$-body, FPf, and FPd models
for the initial conditions of King models with $W_0=3$.}\label{tab:W3}
\begin{tabular}{@{}rcccc}
\hline
$N$  & $t_{\rm rh,i}$ & $T_{\rm half}$ & $T_{\rm half}$ & $T_{\rm half}$  \\
  & & ($N$-body) & (FPf) & (FPd) \\
\hline
128    & $5.13\times 10^0$ & $8.94\times10^1$  & $8.37\times10^1$  & $6.76\times10^1$ \\
256    & $8.13\times 10^0$ & $1.27\times10^2$  & $1.21\times10^2$  & $9.93\times10^1$ \\
512    & $1.35\times 10^1$ & $1.83\times10^2$  & $1.71\times10^2$  & $1.50\times10^2$ \\
1024   & $2.30\times 10^1$ & $2.59\times10^2$  & $2.51\times10^2$  & $2.41\times10^2$ \\
2048   & $4.01\times 10^1$ & $3.73\times10^2$  & $3.73\times10^2$  & $4.00\times10^2$ \\
4096   & $7.11\times 10^1$ & $5.58\times10^2$  & $5.56\times10^2$  & $6.86\times10^2$ \\
8192   & $1.28\times 10^2$ & $8.41\times10^2$  & $8.33\times10^2$  & $1.20\times10^3$ \\
16384  & $2.32\times 10^2$ & $1.18\times10^3$  & $1.26\times10^3$  & $2.16\times10^3$ \\
32768  & $4.24\times 10^2$ & $1.96\times10^3$  & $1.92\times10^3$  & $3.92\times10^3$ \\
65536  & $7.82\times 10^2$ & $3.05\times10^3$  & $2.96\times10^3$  & $7.22\times10^3$ \\
131072 & $1.45 \times 10^3$ & ---               & $4.63\times10^3$  & $1.34\times10^4$ \\
262144 & $2.71 \times 10^3$ & ---               & $7.36\times10^3$  & $2.50\times10^4$ \\
524288 & $5.07 \times 10^3$ & ---               & $1.19\times10^4$  & $4.68\times10^4$ \\
1048576& $9.54 \times 10^3$ & ---               & $1.97\times10^4$  & $8.82\times10^4$ \\
2097152& $1.80 \times 10^4$ & ---               & $3.33\times10^4$  & $1.67\times10^5$ \\
\hline
\end{tabular}
\end{table}

Fig.~\ref{fig:thalfW3} shows the half-mass time $T_{\rm half}$,
which is the time required for a cluster to lose a half of its initial mass,
as a function of the initial half-mass relaxation time $t_{\rm rh,i}$.
Here the half-mass relaxation time is defined by
\begin{equation}
t_{\rm rh} = 0.138\frac{N^{1/2}r_{\rm h}^{3/2}}
              {G^{1/2}  m^{1/2}  \ln (\gamma N)} \label{eq:trh}
\end{equation}
(\citealt{s87}, Chapt. 2) with $\gamma=0.11$ \citep{gh94a}.
The results are summarised also in Table~\ref{tab:W3}.
The FPf and $N$-body models show good agreement over the whole range of $N$
where the comparison is made.
The scaling $T_{\rm half} \propto t_{\rm rh}^{3/4}$ 
gives a reasonable fit to the results of these models
as \citet{b01} found.

The results of FPd models are also shown in Fig.~\ref{fig:thalfW3}.
In these models the parameter $\nu_{\rm e}=2.5$ is used for equation~(\ref{eq:lo}) \citep{tpz00}.
The FPd models show clearly a different scaling from the other models; 
$T_{\rm half} \propto t_{\rm rh}$ expect for models with very short $t_{\rm rh}$ (i.e. small $N$ ).

\begin{figure}
\begin{center}
\includegraphics[width=84mm]{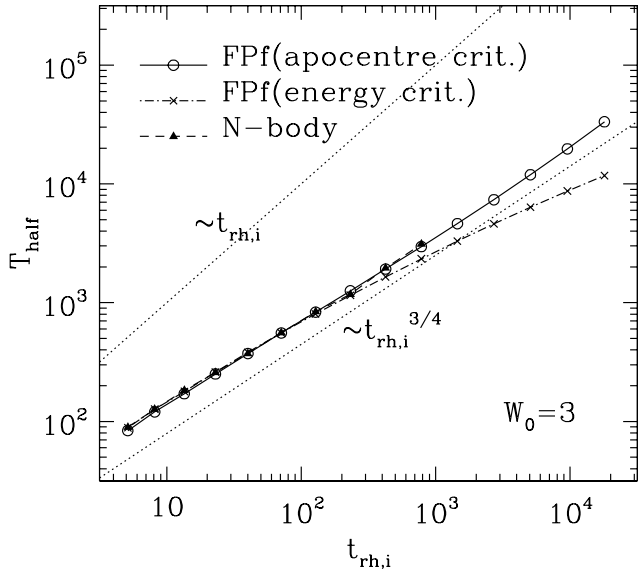}
\end{center}
\caption{Same as Fig.~\ref{fig:thalfW3}, 
but FPf models with the energy criterion are compared with those with the apocentre criterion and
the $N$-body models.
}\label{fig:thalfW3e}
\end{figure}

In Fig.~\ref{fig:thalfW3e} FPf models with the energy criterion are compared with
those with the apocentre criterion as well as the $N$-body models.
We have set $\nu_{\rm e}=5$ in the energy-criterion models so that their mass evolution
reasonably agrees with that of the $N$-body models for small $N$.
There is no significant difference between the energy-criterion models and the other models
for $t_{\rm rh,i} \la 100$,
but the energy-criterion models tend to lose mass much faster as $t_{\rm rh,i}$ increases.
This indicates that the apocentre criterion is a better escape criterion for FPf models.

\begin{figure}
\begin{center}
\includegraphics[width=84mm]{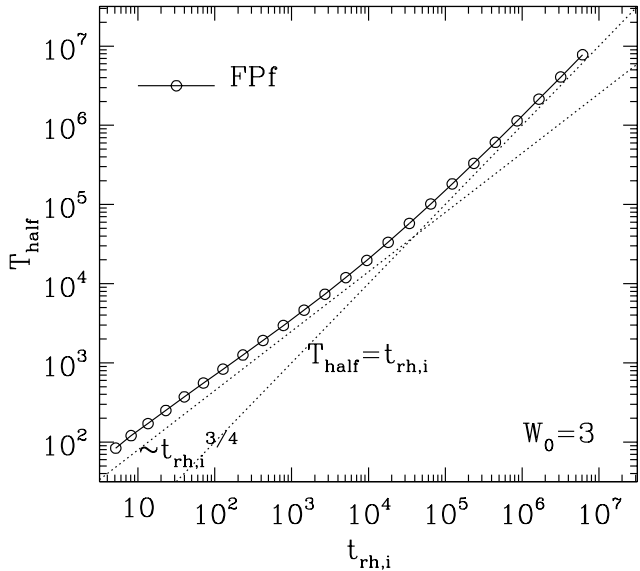}
\end{center}
\caption{Same as Fig.~\ref{fig:thalfW3}, but 
FPf models with $N$ up to $2^{30}$ are shown.
The steeper dotted line represents the relation $T_{\rm half} = t_{\rm rh,i}$.
}\label{fig:thalfW3ex}
\end{figure}

\begin{figure}
\begin{center}
\includegraphics[width=84mm]{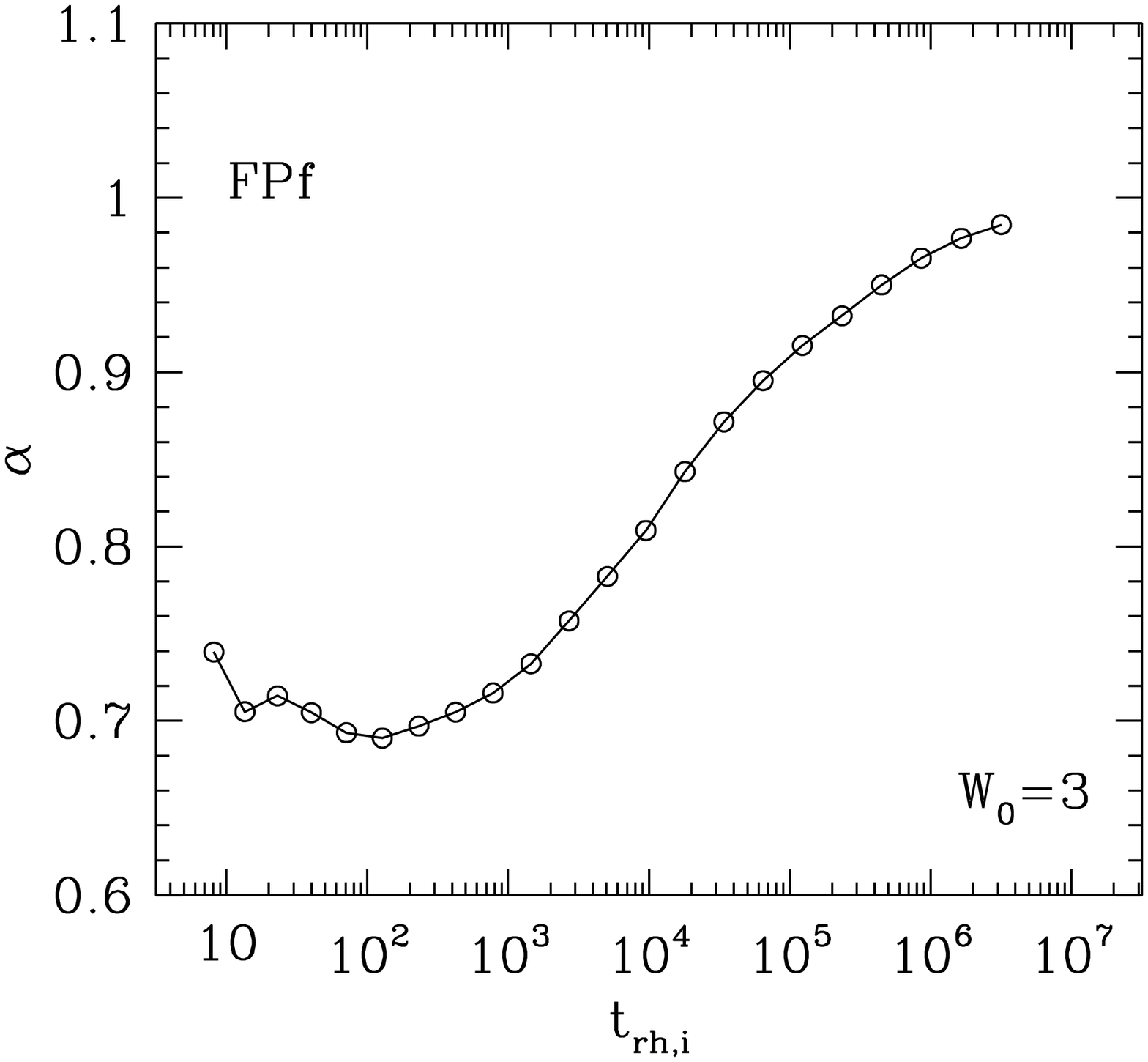}
\end{center}
\caption{Logarithmic slope $\alpha = d\log T_{\rm half}/d\log t_{\rm rh,i}$ 
as a function of the initial half-mass relaxation time $t_{\rm rh,i}$
for the models shown in Fig.~\ref{fig:thalfW3ex}.}
\label{fig:thalfW3expw}
\end{figure}

\begin{table}
\caption{Half-mass times $T_{\rm half}$ given by FPf models
for the initial conditions of $W_0=3$ King models with very large $N$.}\label{tab:W3ex}
\begin{tabular}{@{}cccc}
\hline
$N$  & $t_{\rm rh,i}$ & $T_{\rm half}$ & $T_{\rm half}/t_{\rm rh,i}$\\
\hline
$2^{22}\ (\approx 4.19 \times 10^6)$ & $3.41 \times 10^4$ & $5.77 \times 10^4$  & 1.69 \\
$2^{23}\ (\approx 8.39 \times 10^6)$ & $6.47 \times 10^4$ & $1.02 \times 10^5$  & 1.57 \\
$2^{24}\ (\approx 1.68 \times 10^7)$ & $1.23 \times 10^5$ & $1.82 \times 10^5$  & 1.48 \\
$2^{25}\ (\approx 3.36 \times 10^7)$ & $2.35 \times 10^5$ & $3.31 \times 10^5$  & 1.41 \\
$2^{26}\ (\approx 6.71 \times 10^7)$ & $4.50 \times 10^5$ & $6.09 \times 10^5$  & 1.35\\
$2^{27}\ (\approx 1.34 \times 10^8)$ & $8.62 \times 10^5$ & $1.14 \times 10^6$  & 1.32 \\
$2^{28}\ (\approx 2.68 \times 10^8)$ & $1.65 \times 10^6$ & $2.14 \times 10^6$  & 1.30 \\
$2^{29}\ (\approx 5.37 \times 10^8)$ & $3.18 \times 10^6$ & $4.07 \times 10^6$  & 1.28 \\
$2^{30}\ (\approx 1.07 \times 10^9)$ & $6.12 \times 10^6$ & $7.77 \times 10^6$  & 1.27 \\
\hline
\end{tabular}
\end{table}

As stated above, 
the results of the FPf models shown in Fig.~\ref{fig:thalfW3} are reasonably well
described by the scaling law $T_{\rm half} \propto t_{\rm rh,i}^{3/4}$.
However we should not expect this scaling continues to hold in the limit of large $N$.
If this scaling continues, the half-mass time measured in the units of the half-mass relaxation time,
$T_{\rm half}/t_{\rm rh,i}$, would go to zero as $N \to \infty$.
This must be impossible because the mass loss is driven by two-body relaxation.
In order to see the scaling of $T_{\rm half}$ in the limit of large $N$,
we have calculated FPf models with very large $N$, 
$N=2^{22}\approx 4.19\times 10^6$ to $2^{30}\approx 1.07\times 10^9$,
which are much larger than typical numbers of stars in globular clusters.
The results of these models are shown in Table~\ref{tab:W3ex} and Fig.~\ref{fig:thalfW3ex}.
In Fig.~\ref{fig:thalfW3ex} we see that $T_{\rm half}$ is nearly proportional to $t_{\rm rh,i}$
for very large $N$ clusters, say, for $t_{\rm rh,i} \ga 10^5$ or $N \ga 10^7$.
This trend is more qualitatively shown in Fig.~\ref{fig:thalfW3expw},
where the change in the logarithmic slope,
\begin{equation}
\alpha = \frac{d \log T_{\rm half}}{d \log t_{\rm rh,i}}, \label{eq:alpha}
\end{equation}
is plotted.
The slope $\alpha$ approaches one as $N$ increases.
The ratio $T_{\rm half}/t_{\rm rh,i} \approx 1.3$ for our largest-$N$ models.

\begin{figure}
\begin{center}
\includegraphics[width=84mm]{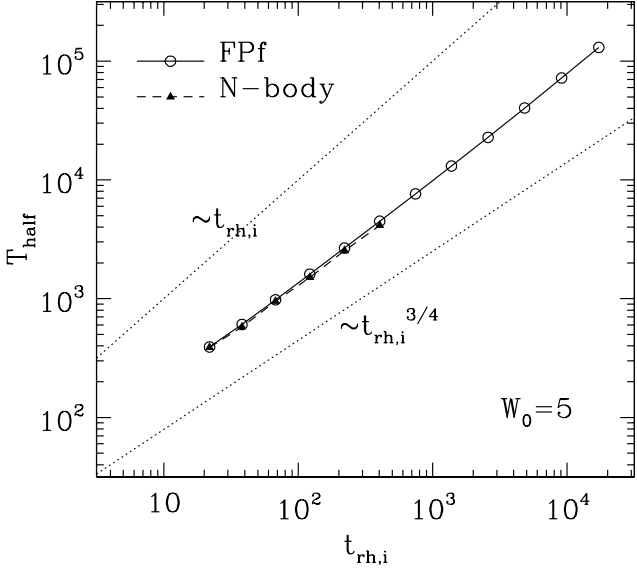}
\end{center}
\caption{Same as Fig.~\ref{fig:thalfW3}, but for the initial conditions of $W_0=5$ King models.
FPf models with the apocentre criterion and $N$-body models are shown.
}\label{fig:thalfW5}
\end{figure}

\begin{table}
\caption{Half-mass times $T_{\rm half}$ given by $N$-body and FPf models
for the initial conditions of King models with $W_0=5$.}\label{tab:W5}
\begin{tabular}{@{}rccc}
\hline
$N$ & $t_{\rm rh,i}$ & $T_{\rm half}$ & $T_{\rm half}$  \\
 & & ($N$-body) & (FPf) \\
\hline
1024      & $2.19 \times 10^1$ & $3.89\times10^2$ & $3.92\times10^2$ \\
2048      & $3.82 \times 10^1$ & $5.78\times10^2$ & $6.07\times10^2$ \\
4096      & $6.77 \times 10^1$ & $9.51\times10^2$ & $9.77\times10^2$ \\
8192      & $1.22 \times 10^2$ & $1.51\times10^3$ & $1.61\times10^3$ \\
16384     & $2.21 \times 10^2$ & $2.54\times10^3$ & $2.67\times10^3$ \\
32768     & $4.04 \times 10^2$ & $4.14\times10^3$ & $4.49\times10^3$ \\
65536     & $7.45 \times 10^2$ & ---              & $7.62\times10^3$ \\
131072    & $1.38 \times 10^3$ & ---              & $1.31\times10^4$ \\
262144    & $2.58 \times 10^3$ & ---              & $2.28\times10^4$ \\
524288    & $4.83 \times 10^3$ & ---              & $4.03\times10^4$ \\
1048576   & $9.09 \times 10^3$ & ---              & $7.20\times10^4$ \\
2097152   & $1.72 \times 10^4$ & ---              & $1.30\times10^5$ \\
\hline             
\end{tabular}
\end{table}

We have performed simulations also for the initial conditions of $W_0=5$ King models.
The half-mass times of $N$-body and FPf models for $W_0=5$ are summarised in
Table~\ref{tab:W5} and are plotted in Fig.~\ref{fig:thalfW5}.
Here we find good agreement again.
The same parameter $\nu_{\rm e}=7$ is used for both the $W_0=3$ and $W_0=5$ clusters.
In Fig.~\ref{fig:thalfW5} the slope of the $\log t_{\rm rh,i}$--$\log T_{\rm half}$ relation seems to be in between 
$3/4$ and 1.
This point is further examined in subsection \ref{ssec:field_strength}.

\subsection{Dependence on the escape-time function} \label{ssec:escape_time}

\citet{b01} argued that the scaling $T_{\rm half} \propto t_{\rm rh}^{3/4}$ 
can be explained by a steady state solution of a simple model 
for the evolution of potential escapers (see equation (12) of his paper).
His model adopts the escape time-scale $t_{\rm e}$ of equation (\ref{eq:fhte}).
If a different function is assumed for $t_{\rm e}$, his model predicts
a different scaling law.
It is shown that the scaling
\begin{equation}
T_{\rm half} \propto t_{\rm rh}^\frac{\beta+1}{\beta+2} \label{eq:thbeta}
\end{equation}
is obtained for $t_{\rm e} \propto (E-E_{\rm crit})^{-\beta}$ (see Appendix A).
It is interesting to see if this prediction is confirmed by the results of our FPf models.

\begin{figure}
\begin{center}
\includegraphics[width=84mm]{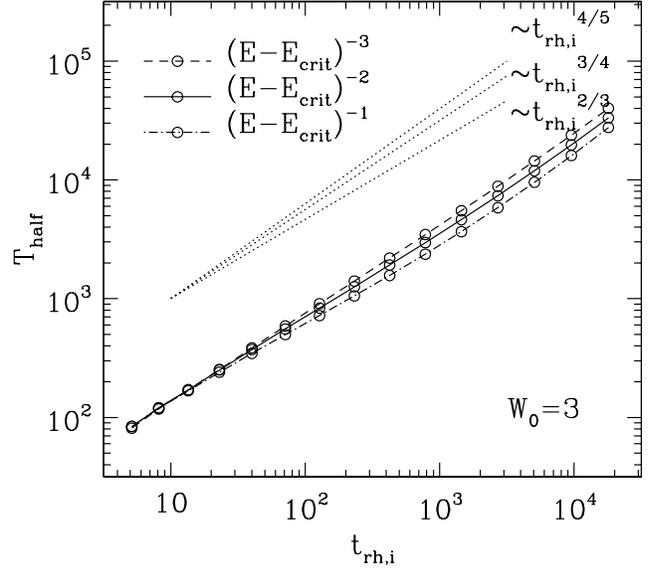}
\end{center}
\caption{Same as Fig.~\ref{fig:thalfW3}, but FPf models with 
different functional forms of $t_{\rm e}(E) \propto (E-E_{\rm crit})^{-\beta} $ ($\beta=1, 2, 3$) are compared. 
The dotted lines represent scalings $t_{\rm rh}^{2/3}$, $t_{\rm rh}^{3/4}$ and $t_{\rm rh}^{4/5}$,
which are predicted by the simple steady-solution model 
for $\beta=1$, 2 and 3, respectively (see text).
}\label{fig:thalfW3p}
\end{figure}

We have performed FP runs using a generalized form of equation~(\ref{eq:te}),
\begin{equation}
\frac{1}{t_{\rm e}(E)} = 
  \frac{\nu_{\rm e}}{t_{\rm tid}} \left( 1-\frac{E}{E_{\rm crit}} \right)^\beta, \label{eq:tebeta}
\end{equation}
with $\beta=1$ and 3.
Fig.~\ref{fig:thalfW3p} plots the half-mass time against the initial half-mass relaxation time
for these runs as well as for the standard runs, where King models with $W_0=3$ are used as initial conditions.
The value of $\nu_{\rm e}$ has been adjusted 
so that the non-standard models should have roughly the same half-mass times
with those of the standard ones for lower $N$;
$\nu_{\rm e}=7/3$ and $7\times 3$ for $\beta=1$ and 3, respectively.

The results of the FPf models actually depend on $\beta$, 
but the degree of the dependence is weaker than predicted by equation~(\ref{eq:thbeta}).
While this equation predicts the slopes 2/3, 3/4 and 4/5 for $\beta=1$, 2 and 3, respectively,
linear least-squares fitting of the data in Fig.~\ref{fig:thalfW3p}
gives the slopes 0.69, 0.72 and 0.75.
When the fitting is done only for $N \ge 16384$, the slopes are 0.75, 0.75 and 0.77.
Thus the scaling law $T_{\rm half} \propto t_{\rm rh}^{3/4}$
is not a bad approximation in all the cases investigated here.
This is not consistent with equation~(\ref{eq:thbeta}).

\subsection{Dependence on the strength of the tidal field} \label{ssec:field_strength}

\citet{tf05} found that the dependence of $T_{\rm half}$ on $t_{\rm rh,i}$ is affected by the strength
of the tidal field and that the logarithmic slope $\alpha$, defined by equation~(\ref{eq:alpha}),
approaches unity as the strength of the tidal field decreases.
In order to confirm their findings,
we have calculated FPf models for the initial conditions where the initial tidal radius $r_{\rm t,i}$
is greater than the King cutoff radius $r_{\rm K}$ (i.e. the radius at which the density drops to zero)
for each value of $W_0$.
On the other hand, all the models presented above are calculated for the initial conditions with 
$r_{\rm t,i}=r_{\rm K}$.

\begin{figure}
\begin{center}
\includegraphics[width=84mm]{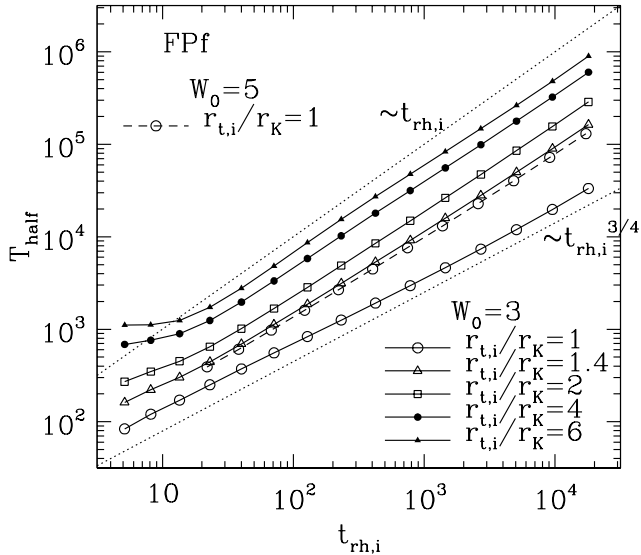}
\end{center}
\caption{Same as Fig.~\ref{fig:thalfW3}, but FPf models for the initial conditions 
of King models with $r_{\rm t,i} > r_{\rm K}$ are compared with
the cases of  $r_{\rm t,i} = r_{\rm K}$.
}\label{fig:thalfW3rk}
\end{figure}

\begin{figure}
\begin{center}
\includegraphics[width=84mm]{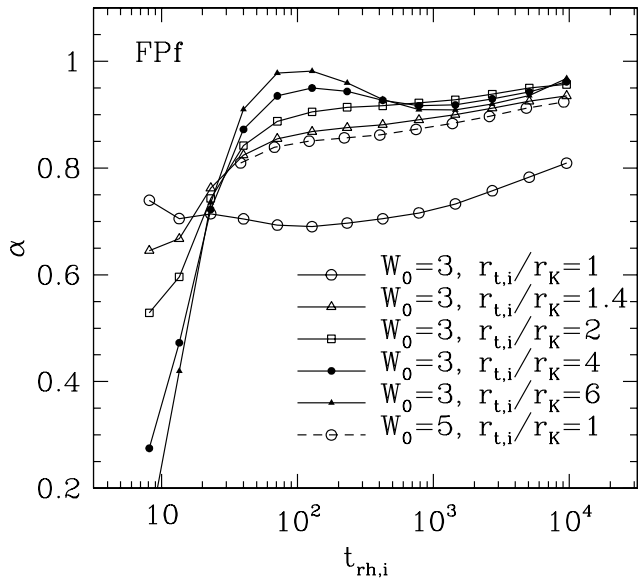}
\end{center}
\caption{Logarithmic slope $\alpha = d\log T_{\rm half}/d\log t_{\rm rh,i}$ 
as a function of the initial half-mass relaxation time $t_{\rm rh,i}$.
The models are the same as those shown in Fig.~\ref{fig:thalfW3rk}.
}\label{fig:thalfpw}
\end{figure}

Table~\ref{tab:W3rk} lists the half-mass times 
for $W_0=3$ King models with $r_{\rm t,i}/r_{\rm K}=1.4$, 2, 4 and 6,
and Fig.~\ref{fig:thalfW3rk} illustrates these results.
In this figure the results for $W_0=3$ and $W_0=5$ King models with $r_{\rm t,i}/r_{\rm K}=1$
are also plotted.
Note that the ratio $r_{\rm K}(W_0=5)/r_{\rm K}(W_0=3) \approx 1.4$.
Fig.~\ref{fig:thalfpw} shows the variation of $\alpha$ with $t_{\rm rh,i}$.

The results shown in Figs.~\ref{fig:thalfW3rk} and \ref{fig:thalfpw} confirm the findings of \citet{tf05}.
The dependence of $T_{\rm half}$ on $t_{\rm rh,i}$ does depend on the strength of the tidal field.
In the limit of $r_{\rm t,i}/r_{\rm K} \to \infty$ and $N \to \infty$,
it is expected that $\alpha \to 1$.

Note that the curve for $W_0=3$ King models with $r_{\rm t,i}/r_{\rm K}=1.4$ 
lies very close to that for $W_0=5$ King models with $r_{\rm t,i}/r_{\rm K}=1$
in each of Figs.~\ref{fig:thalfW3rk} and \ref{fig:thalfpw}.
This indicates that the mass-loss time-scale 
does not depend very much on the initial concentration of the cluster
but is mainly determined by the strength of the tidal field,
as was found by \citet{tf05}.

\begin{table*}
\begin{minipage}{105mm}
\caption{Half-mass times $T_{\rm half}$ given by FPf models
for the initial conditions of King models with $W_0=3$
and $r_{\rm t,i} > r_{\rm K}$.}\label{tab:W3rk}
\begin{tabular}{@{}rccccc}
\hline
$N$ & $T_{\rm half}$  & $T_{\rm half}$ & $T_{\rm half}$ & $T_{\rm half}$ \\
    & ($r_{\rm t,i}/r_{\rm K}=1.4$) & ($r_{\rm t,i}/r_{\rm K}=2$) & ($r_{\rm t,i}/r_{\rm K}=4$) & ($r_{\rm t,i}/r_{\rm K}=6$) \\
\hline
128 & $1.62\times10^2$ & $2.71\times10^2$ & $6.86\times10^2$ & $1.11\times10^3$ \\
256 & $2.22\times10^2$ & $3.48\times10^2$ & $7.61\times10^2$ & $1.12\times10^3$ \\
512 & $3.02\times10^2$ & $4.52\times10^2$ & $8.95\times10^2$ & $1.25\times10^3$ \\
1024 & $4.44\times10^2$ & $6.48\times10^2$ & $1.24\times10^3$ & $1.73\times10^3$ \\
2048 & $6.93\times10^2$ & $1.01\times10^3$ & $1.97\times10^3$ & $2.78\times10^3$ \\
4096 & $1.12\times10^3$ & $1.68\times10^3$ & $3.33\times10^3$ & $4.83\times10^3$ \\
8192 & $1.87\times10^3$ & $2.84\times10^3$ & $5.82\times10^3$ & $8.64\times10^3$ \\
16384 & $3.14\times10^3$ & $4.89\times10^3$ & $1.02\times10^4$ & $1.54\times10^4$ \\
32768 & $5.33\times10^3$ & $8.49\times10^3$ & $1.80\times10^4$ & $2.73\times10^4$ \\
65536 & $9.16\times10^3$ & $1.49\times10^4$ & $3.16\times10^4$ & $4.77\times10^4$ \\
131072 & $1.59\times10^4$ & $2.64\times10^4$ & $5.56\times10^4$ & $8.34\times10^4$ \\
262144 & $2.80\times10^4$ & $4.72\times10^4$ & $9.89\times10^4$ & $1.48\times10^5$ \\
524288 & $4.99\times10^4$ & $8.54\times10^4$ & $1.78\times10^5$ & $2.64\times10^5$ \\
1048576 & $8.98\times10^4$ & $1.56\times10^5$ & $3.24 \times 10^5$ & $4.79 \times 10^5$ \\
2097152 & $1.63\times10^5$ & $2.87 \times 10^5$ & $6.02 \times 10^5$ & $8.99 \times 10^5$ \\
\hline             
\end{tabular}
\end{minipage}
\end{table*}

\subsection{Comparison with $N$-body models: multi-mass clusters} \label{ssec:multi-mass}

So far we have concentrated on single-mass clusters.
Here we consider the evolution of multi-mass clusters
comparing our FP models with the $N$-body models of \citet{gb08}.
They performed $N$-body simulations of clusters on circular orbits around a point-mass galaxy.
In their simulations the initial mass function (IMF) is given by $dN/dm \propto m^{-2.35}$ 
with the ratio $m_{\rm max}/m_{\rm min}=30$.
Stellar evolution is not considered in their simulations.
The clusters initially have the density distribution of King models with $W_0=5$.
The ratio of the initial tidal radius to the King radius $r_{\rm t,i}/r_{\rm K}$ is varied from 1 to 8.
The results of the simulations of \citet{gb08} are summarised in their Table~1.
Note that they use different notations from ours: $r_{\rm J}$ is for the tidal (Jacobi) radius
 and $r_{\rm t}$ is for the King radius.

\begin{table}
\caption{Half-mass times $T_{\rm half}$ given by $N$-body \citep{gb08} and FPf models
for the initial conditions of multi-mass King models with $W_0=5$ and $r_{\rm t,i}/r_{\rm K}=1$.
Three sets of the parameters $(\gamma, \nu_{\rm e})$ are used for the FPf models.}
\label{tab:mm_gamma}
\begin{tabular}{@{}rcccc}
\hline
$N$ & $T_{\rm half}$ & $T_{\rm half}$  & $T_{\rm half}$ & $T_{\rm half}$  \\
 & ($N$-body) & (FPf)  & (FPf) & (FPf) \\
 &            & (0.11, 7)  & (0.02, 7) & (0.02, 40) \\
\hline
1024      & $1.14 \times 10^2$ & $1.20 \times 10^2$ & $1.69 \times 10^2$  & $1.17 \times 10^2$ \\
2048      & $1.74 \times 10^2$ & $1.87 \times 10^2$ & $2.54 \times 10^2$  & $1.80 \times 10^2$ \\
4096      & $2.69 \times 10^2$ & $2.86 \times 10^2$ & $3.75 \times 10^2$  & $2.72 \times 10^2$ \\
8192      & $4.35 \times 10^2$ & $4.39 \times 10^2$ & $5.59 \times 10^2$  & $4.18 \times 10^2$ \\
16384     & $6.70 \times 10^2$ & $6.90 \times 10^2$ & $8.57 \times 10^2$  & $6.59 \times 10^2$ \\
32768     & $1.06 \times 10^3$ & $1.12 \times 10^3$ & $1.36 \times 10^3$  & $1.08 \times 10^3$ \\
\hline             
\end{tabular}
\end{table}

\begin{figure}
\begin{center}
\includegraphics[width=84mm]{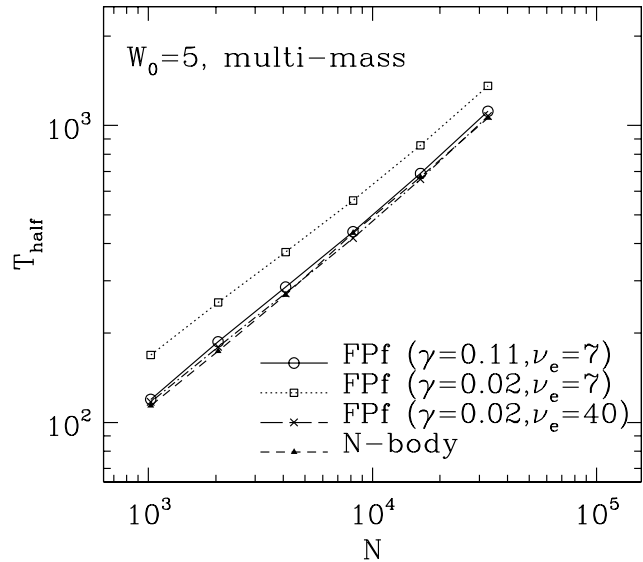}
\end{center}
\caption{
Half-mass time $T_{\rm half}$ as a function of the initial number of stars $N$
for $W_0=5$ King models with the IMF $dN/dm \propto m^{-2.35}$ ($m_{\rm max}/m_{\rm min}=30$).
FPf models with three different sets of the parameters $\gamma$ and $\nu_{\rm e}$ 
are compared with
the $N$-body models of \citet{gb08}.
}
\label{fig:mm_gamma}
\end{figure}

FPf models are calculated for the same initial conditions as those of \citet{gb08}.
The results for $r_{\rm t,i}/r_{\rm K}=1$ 
are summarised in Table~\ref{tab:mm_gamma} and Fig.~\ref{fig:mm_gamma}.
There the results of the FPf models with three different sets of parameters $\gamma$ and $\nu_{\rm e}$ are reported.
\citet{gh94a} estimated the best value of $\gamma=0.11$ for single-mass clusters 
by comparing $N$-body models with FP and gas models. 
Similarly \citet{gh96} obtained $\gamma=0.02$ for multi-mass with
the IMF $dN/dm \propto m^{-2.5} (m_{\rm max}/m_{\rm min}=37.5)$.
We have calculated FPf models for multi-mass clusters using these two values of $\gamma$.

Fig.~\ref{fig:mm_gamma} shows that the parameter set $(\gamma, \nu_{\rm e}) = (0.11,7)$ adopted for single-mass clusters
gives good fit to the $N$-body models also for multi-mass clusters.
On the other hand the parameter set $(\gamma, \nu_{\rm e}) = (0.02,7)$ results in a clear deviation from the $N$-body models.
If we stick to $\gamma=0.02$,
the value of $\nu_{\rm e}$ needs to be increased to about 40 in order to obtain good agreement with the $N$-body models.
We will discuss in more detail what values of the parameters we should choose in the next section.

\begin{table}
\caption{Half-mass times $T_{\rm half}$ given by FPf models
for the initial conditions of multi-mass King models with $W_0=5$ and $r_{\rm t,i}/r_{\rm K}=2$, 4, 8.
The adopted parameter set is $(\gamma, \nu_{\rm e})=(0.11, 7)$.}
\label{tab:mm_rt}
\begin{tabular}{@{}rccc}
\hline
$N$ & $T_{\rm half}$ & $T_{\rm half}$ & $T_{\rm half}$ \\
 & ($r_{\rm t,i}/r_{\rm K}=2$) & ($r_{\rm t,i}/r_{\rm K}=4$) & ($r_{\rm t,i}/r_{\rm K}=8$) \\
\hline
1024      & $3.34 \times 10^2$  & $8.03 \times 10^2$  & $1.81 \times 10^3$  \\
2048      & $5.12 \times 10^2$ & $1.17 \times 10^3$ & $2.41 \times 10^3$  \\
4096      & $7.79 \times 10^2$  & $1.72 \times 10^3$ & $3.35 \times 10^3$ \\
8192      & $1.21 \times 10^3$ & $2.65 \times 10^3$ & $5.08 \times 10^3$  \\
16384     & $1.95 \times 10^3$ & $4.30 \times 10^3$ & $8.33 \times 10^3$  \\
32768     & $3.26 \times 10^3$ & $7.32 \times 10^3$ & $1.45 \times 10^4$ \\
\hline             
\end{tabular}
\end{table}

\begin{table}
\caption{
Same as Table~\ref{tab:mm_rt}, but
the results of FPf models with the parameter set $(\gamma, \nu_{\rm e})=(0.02, 40)$ are listed.}
\label{tab:mm_rt2}
\begin{tabular}{@{}rccc}
\hline
$N$ & $T_{\rm half}$ & $T_{\rm half}$ & $T_{\rm half}$ \\
 & ($r_{\rm t,i}/r_{\rm K}=2$) & ($r_{\rm t,i}/r_{\rm K}=4$) & ($r_{\rm t,i}/r_{\rm K}=8$) \\
\hline
1024      & $3.68 \times 10^2$ & $9.28 \times 10^2$ & $2.20 \times 10^3$  \\
2048      & $5.53 \times 10^2$ & $1.32 \times 10^3$ & $2.87 \times 10^3$  \\
4096      & $8.25 \times 10^2$ & $1.89 \times 10^3$ & $3.86 \times 10^3$ \\
8192      & $1.26 \times 10^3$ & $2.84 \times 10^3$ & $5.63 \times 10^3$  \\
16384     & $2.01 \times 10^3$ & $4.53 \times 10^3$ & $8.92 \times 10^3$  \\
32768     & $3.55 \times 10^3$ & $7.62 \times 10^3$ & $1.52 \times 10^4$ \\
\hline             
\end{tabular}
\end{table}

\begin{figure}
\begin{center}
\includegraphics[width=84mm]{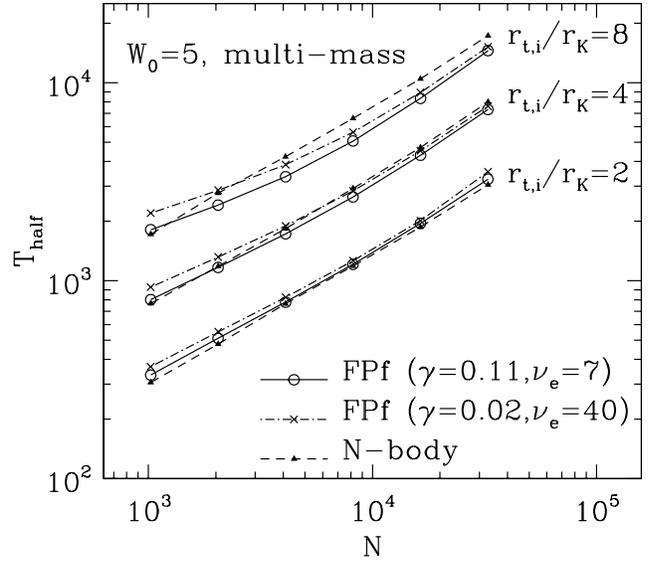}
\end{center}
\caption{
Same as Fig.~\ref{fig:mm_gamma}, but FPf models 
are compared with the $N$-body models of \citet{gb08}
for the clusters with $r_{\rm t,i} > r_{\rm K}$.
The adopted parameter sets for the FPf models are $(\gamma, \nu_{\rm e})=(0.11, 7)$ and $(0.02, 40)$.
}
\label{fig:mm_rt}
\end{figure}

The results for the initial conditions with $r_{\rm t,i}>r_{\rm K}$ 
are shown in Tables~\ref{tab:mm_rt} and \ref{tab:mm_rt2} and Fig.~\ref{fig:mm_rt}.
The results of \citet{gb08} are not shown in these tables (see their Table 1).
Fig.~\ref{fig:mm_rt} shows that the FPf models with $(\gamma, \nu_{\rm e})=(0.11, 7)$
are in good agreement with the $N$-body models
for $r_{\rm t,i}/r_{\rm K}=2$ and 4.
The FPf models with $(\gamma, \nu_{\rm e})=(0.02, 40)$ are a little farther to the $N$-body models
but still follow them rather well.
However, for $r_{\rm t,i}/r_{\rm K}=8$, a noticeable difference is observed between the FPf and $N$-body models;
in Fig.~\ref{fig:mm_rt} the curve for the $N$-body models is approximately linear 
but the slopes of the curves for the FPf models apparently change with $N$.
Neither parameter set reproduces the results of the $N$-body models as well as
in the cases of $r_{\rm t,i}/r_{\rm K}<8$.
The reason for this discrepancy is not clear at present, but
there is a possibility that very early core-collapse in the models with $r_{\rm t,i}/r_{\rm K}=8$ is, 
at least partially, responsible for it.
The FPf model with $(\gamma, \nu_{\rm e})=(0.11, 7)$ and $r_{\rm t,i}/r_{\rm K}=8$
experiences core collapse (bounce) at $t=0.006 T_{\rm half}$ for $N=1024$,
and at $t=0.03 T_{\rm half}$ for $N=32768$.
The Coulomb logarithm may take different values for pre-collapse and post-collapse stages
(see the next section), which affects the time-scale of the evolution of FP models.


\section{Discussion}

We have shown that FP models can well follow the mass evolution of star clusters in a tidal field 
if a new scheme for treating potential escapers is implemented.
This is the first time the effect of re-scattering of potential escapers has been taken into account
in FP models.
Although Takahashi \& Portegies Zwart (1998, 2000) showed that anisotropic FP models are in good agreement
with $N$-body models for the mass evolution of star clusters in a galaxy,
the tidal field is treated as a tidal cutoff rather than an actual force field.
In the present study we have found that our new FP models are in good agreement
with $N$-body models calculated with the inclusion of the tidal force field.
Thus the new scheme has improved the accuracy of FP models.

\citet{b01} argued that some potential escapers are scattered back to lower energies 
before they leave the cluster and that this complicates the scaling of the mass-loss time.
The success of our models is consistent with his argument.
Actually our equation for potential escapers, equation~(\ref{eq:fp_pe}), 
can be regarded as a generalization of the equation of his toy model, his equation (12), 
used for explaining the scaling $T_{\rm half} \propto t_{\rm rh}^{3/4}$.

The toy model of \citet{b01} is useful 
for giving us insight into the effect of potential escapers on the cluster evolution.
On the other hand, the results presented in subsection~\ref{ssec:escape_time}
have revealed the limitation of the model.
When the energy dependence of the escape time is artificially changed from the true one,
the toy model does not correctly explain the results of our FP models.
This failure of the toy model is not a big surprise,
because it is only a simplified model based on many assumptions, some of which are not very realistic.
For example, our simulations show that an exact steady state is never established,
but the toy model assumes a steady state.
In addition, the scaling of the cluster lifetime depends on the strength of the tidal field, 
as found by \citet{tf05} and confirmed by the present study,
but the toy model does not take account of the strength of the tidal field.


Our FP models show good agreement with $N$-body models not only for single-mass clusters
but also for multi-mass clusters.
However, we have encountered a difficulty in determining proper values of the two parameters,
$\gamma$ and $\nu_{\rm e}$, in the FP models.
As shown in subsection~\ref{ssec:multi-mass},
the parameter set $(\gamma, \nu_{\rm e})=(0.11, 7)$ brings good agreement 
for both single-mass and multi-mass clusters.
Since the escape time-scale $t_{\rm e}$ given by equation~(\ref{eq:te}) is expected to be independent of stellar mass,
it is natural that the same value of the parameter $\nu_{\rm e}$ is applicable 
to both single-mass and multi-mass clusters.

On the other hand, the value of $\gamma$ is expected to depend on the stellar mass function.
\citet{h75} argued theoretically that the value of $\gamma$ is generally smaller in multi-mass clusters
than in single-mass clusters.
Based on the results of $N$-body simulations,
\citet{gh94a} obtained a value of $\gamma=0.11$ for isolated single-mass clusters,
and \citet{gh96} obtained a much smaller value, $\gamma=0.02$, for isolated multi-mass clusters
having an IMF similar to the IMF used in our simulations.

When we adopt the value of $\gamma=0.02$ for multi-mass clusters,
we have to use a much larger value of $\nu_{\rm e}$, $\nu_{\rm e}=40$, 
than the best value of $\nu_{\rm e}=7$ for single-mass clusters,
in order to obtain good agreement with $N$-body models.
Thus we have not found a parameter set satisfying both
the independence of $\nu_{\rm e}$ on the mass function and
the dependence of $\gamma$ on it.
It needs further investigation to solve this incompatibility,
but even the determination of $\gamma$ itself is not a simple task.
For example, \citet{gh94b} obtained the best value of $\gamma=0.035$ 
by examining the post-collapse evolution of $N$-body models
of isolated single-mass clusters.
This value is much smaller than the value of $\gamma=0.11$ obtained for 
pre-collapse single-mass clusters.
These results suggest that the value of $\gamma$
changes along with the evolution of clusters.
It may also change with radius within a cluster \citep{gh94a}.


\citet{fh00} theoretically estimated not only the energy dependence of the escape time-scale $t_{\rm e}$ 
but also its numerical coefficient, which is given in their equation (9).
If we ignore the difference between energy $E$ and the Jacobi integral $E_{\rm J}$,
their estimate for a $W_0=3$ King model leads to a value of $\nu_{\rm e}=29$.
This is about four times larger than our best value of $\nu_{\rm e}=7$ for single-mass clusters.
However,
\citet{fh00} also did numerical experiments and found that their theoretical estimate of $t_{\rm e}$ is too small; 
escape time-scales obtained from the numerical experiments are more than a few times larger than the theoretical one.
Therefore our value $\nu_{\rm e}=7$ is not inconsistent with the result of \citet{fh00}.
On the other hand,
our value of $\nu_{\rm e}=40$ for multi-mass clusters with $\gamma=0.02$
is a little larger than their theoretical estimate.


Another issue not addressed in the present paper is how the mass profile of the parent galaxy
affects the results.
In all the simulations presented here
we assume that the parent galaxy is represented by a point mass.
On the other hand, \citet{tf10} showed that the mass-loss time-scale depends on
the mass profile of the parent galaxy;
the time-scale increases as the mass profile gets shallower.
Therefore we expect that the parameter $\nu_{\rm e}$ depends on the mass profile of the parent galaxy.
This issue will be examined in a future study.

\section{Conclusion}

In this paper we have developed new FP models of globular clusters in a steady galactic tidal field.
Our FP models are novel in the method of treating escapers:
potential escapers are allowed to experience gravitational scattering with other stars 
before they really leave clusters.
The new method has been devised in order to construct more realistic models of star clusters in a tidal field
compared to simple tidal-cutoff models as in previous studies.
The mass evolution of clusters in a tidal field does not simply scale with the relaxation time,
and our FP models are in good agreement with $N$-body models in this respect.

Our FP models include two parameters $\gamma$ and $\nu_{\rm e}$;
$\gamma$ is the numerical factor in the Coulomb logarithm $\ln (\gamma N)$ and 
$\nu_{\rm e}$ adjusts the speed of the tidal mass loss.
We have determined the best values of $\nu_{\rm e}$ for given values of $\gamma$
by comparing FP results with $N$-body results.
For single-mass clusters the best parameter set is $(\gamma, \nu_{\rm e})=(0.11, 7)$.
This parameter set is applicable to multi-mass clusters as well,
but another set $(\gamma, \nu_{\rm e})=(0.02, 40)$ does work equally well as long as multi-mass clusters are concerned.
The parameter $\nu_{\rm e}$ is expected to depend on the mass profile of the parent galaxy,
though a point-mass galaxy is assumed in all the simulations of the present paper.
Further investigation is required for the determination of the best values of the parameters
$\gamma$ and $\nu_{\rm e}$ under various conditions.

While FP models are generally thought to be less faithful models of globular clusters than $N$-body models,
the present study has significantly improved the accuracy of FP models.
An advantage of FP models is that they can be calculated much faster than $N$-body models.
Therefore FP models are particularly useful when we need to calculate a huge number of models.
For example, when we try to specify the initial conditions of individual clusters,
we have to perform simulations for many sets of the initial conditions,
because the parameter space to be searched is very large.
We believe that our FP models is quite useful for such searching.

\section*{Acknowledgments}

Part of the work was done while the authors visited the Center for
Planetary Science (CPS) in Kobe, Japan, during a visit that was
funded by the HPCI Strategic Program of MEXT.
We are grateful for their hospitality.
HB acknowledges support by the Australian Research Council (ARC) through Future Fellowship Grant FT0991052.
The numerical calculations of the Fokker-Planck models were carried out 
on Altix3700 and SR16000 at YITP in Kyoto University.


\appendix

\section{Estimation of the scaling of the cluster lifetime}\label{sec:ap1}

We follow the arguments given by \citet{b01} and \citet{h01}
in order to derive the scaling law of equation (\ref{eq:thbeta}).

Let $\hat{E}=(E-E_{\rm crit})/|E_{\rm crit}|$
and
assume that the escape time-scale $t_{\rm e}$ has energy-dependence such as
\begin{equation}
t_{\rm e}(\hat{E}) = t_{\rm esc} \hat{E}^{-\beta} \quad(\beta >0) .
\end{equation}
Then Baumgardt's toy model is modified as 
\begin{equation}
\frac{\partial n}{\partial t} 
= \frac{k_1}{t_{\rm rh}} \frac{\partial^2 n}{\partial \hat{E}^2} - \hat{E}^\beta \frac{n}{t_{\rm esc}}, 
\label{eq:toy}
\end{equation}
where $n(\hat{E},t)d\hat{E}$ is the number of stars with energies in the range $(\hat{E}, \hat{E}+d\hat{E})$
and $k_1$ is a constant.
If we assume that the distribution of escapers is nearly in equilibrium,
equation (\ref{eq:toy}) shows that the width of the distribution is approximately given by
\begin{equation}
\Delta \hat{E} \sim \left( \frac{t_{\rm esc}}{t_{\rm rh}} \right)^{\frac{1}{\beta+2}} ,
\end{equation}
and the number of escapers $N_{\rm esc} \sim N \Delta \hat{E}$.
The escape rate $\dot{N}_{\rm esc}$ is estimated to be
\begin{equation}
\dot{N}_{\rm esc} \sim \frac{N_{\rm esc}}{t_{\rm e}(\Delta \hat{E})}
\sim \frac{N}{t_{\rm esc}} \left( \frac{t_{\rm esc}}{t_{\rm rh}} \right)^{\frac{\beta+1}{\beta+2}} .
\end{equation}
Therefore the scaling of the half-mass time is given by
\begin{equation}
T_{\rm half} \sim \frac{N}{\dot{N}_{\rm esc}} \sim t_{\rm rh}^\frac{\beta+1}{\beta+2} t_{\rm esc}^\frac{1}{\beta+2} .
\end{equation}

\label{lastpage}

\end{document}